\documentclass[preprint,prd,showpacs,showkeys]{revtex4-1}
\usepackage{amsmath}
\usepackage{amssymb}
\usepackage{amsfonts}
\usepackage{graphicx}
\usepackage{float}
\usepackage{appendix}

\begin{document}

\title{Ghost Loops are Indispensable in Unitary Gauge}
\author{Er-Cheng Tsai}
\email{ectsai@ntu.edu.tw}
\affiliation{Physics Department, National Taiwan University, Taipei, Taiwan}

\begin{abstract}
It is conventionally taken for granted that the unitary gauge formulation of
quantum gauge field theory has the advantage of preservation unitarity because
only physical fields are involved but has the disadvantage of losing
renormalizability because of severe ultraviolet divergences due to vector
meson propagators. In this paper, we show how to handle the ultraviolet
divergent loops so that the physical amplitudes remain gauge invariant. One of
the consequences we arrive at is that ghost loops are needed to cancel the
divergences due to\ vector mesons and to give gauge invariant physical amplitudes.
\end{abstract}

\pacs{11.10.Kk, 11.15.-q, 11.15.Bt}
\keywords{quantum field; unitary gauge; dimensional regularization }
\maketitle
\affiliation{Physics Department, National Taiwan University, Taipei, Taiwan}

\section{Introduction}

In the unitary gauge, only physical fields are involved but the longitudinal
propagator for the vector meson, $\frac{i}{\left(  k^{2}-M^{2}\right)  }%
\frac{k^{\mu}k^{\nu}}{M^{2}}$, does not vanish like $\frac{1}{k^{2}}$ when
$k\rightarrow\infty$ and hence increases the power count by two and may render
the unitary gauge theory unrenormalizable according to power counting. We show
in this paper that the divergent terms due to longitudinal vector propagators
and ghost loops in the unitary gauge actually cancel themselves under
dimensional regularization. Furthermore, the Feynman integrand for ghost loop
turns out to be loop-momentum independent.

A Ward-Takahashi identity \cite{WTI} involving divergent amplitudes is not
meaningful unless these amplitudes have been regularized. Take, for example,
the well-known identity $k_{\mu}\Pi^{\left(  1\right)  \mu\nu}=0$ for the
one-loop photon self-energy correction $\Pi^{\left(  1\right)  \mu\nu}$ in
QED. The function $\Pi^{\left(  1\right)  \mu\nu}$ may be formally written as%
\[
\Pi^{\left(  1\right)  \mu\nu}=e^{2}\int\frac{d^{4}\ell}{\left(  2\pi\right)
^{4}}tr\left(  \frac{1}{\not \ell -m}\gamma^{\mu}\frac{1}{\not \ell +\not k
-m}\gamma^{\nu}\right)
\]
where $k$ is the external momentum. The identity%
\begin{equation}
\frac{1}{\not \ell -m}\not k  \frac{1}{\not \ell +\not k  -m}=\frac{1}{\ell
-m}-\frac{1}{\not \ell +\not k  -m} \label{qedwti0}%
\end{equation}
allows us to express $k_{\mu}\Pi^{\left(  1\right)  \mu\nu}$ as%
\[
k_{\mu}\Pi^{\left(  1\right)  \mu\nu}=e^{2}\int\frac{d^{4}\ell}{\left(
2\pi\right)  ^{4}}tr\left(  \frac{1}{\ell-m}\gamma^{\nu}-\frac{1}%
{\not \ell +\not k  -m}\gamma^{\nu}\right)  .
\]
The above integral is divergent and hence meaningless. As we know, shifting
loop variables is an operation that is not always legitimate for a divergent
integral and the identity $k_{\mu}\Pi^{\left(  1\right)  \mu\nu}=0$ is merely formal.

Dimensional regularization \cite{HV} ensures that the amplitude $\Pi^{\left(
1\right)  \mu\nu}$ is well-defined and loop momentum shifting is allowed. As a
result, the difference of two terms related by a shift of loop momentum
variable is equal to zero. Thus the dimensionally regularized amplitude
$\Pi^{\left(  1\right)  \mu\nu}$ satisfies $k_{\mu}\Pi^{\left(  1\right)
\mu\nu}=0$. This is the advantage of dimensional regularization. We want to
emphasize that Ward identities can be shown to be satisfied without actually
carrying out the loop integrals, if terms with integrands related by shift of
loop momenta are treated as equal. \ In this paper, we use the 1-loop
self-energy correction of\ the physical Higgs field for the Abelian gauge
theory in both the Feynman and unitary gauges as an example to illustrate how
the extra divergent terms in the unitary gauge cancel themselves among
different diagrams that also include the ghost loops whose integrands turn to
be independent of the loop momentum. Ghost loops were erroneously neglected in
the conventional treatment of unitary gauge.

\section{Abelian Gauge Theory with Higgs}

The Lagrangian for Abelian-Higgs theory \cite{HIGGS} is
\begin{equation}
L=-{\frac{1}{4}}F_{\mu\nu}F^{\mu\nu}+\left(  D^{\mu}\phi\right)  ^{\dagger
}\left(  D_{\mu}\phi\right)  -\frac{1}{2}\lambda g^{2}\left(  \phi^{\dagger
}\phi-\frac{1}{2}\upsilon^{2}\right)  ^{2} \label{e2-1}%
\end{equation}
where%
\begin{align*}
F_{\mu\nu}  &  \equiv\partial_{\mu}A_{\nu}-\partial_{\nu}A_{\mu},\\
D_{\mu}\phi &  \equiv\left(  \partial_{\mu}+igA_{\mu}\right)  \phi
\end{align*}
Define two Hermitian fields $H$ and $\phi_{2}$ for the real and imaginary
parts of the complex scalar field by%
\[
\phi=\frac{1}{\sqrt{2}}\left(  H+\upsilon+i\phi_{2}\right)
\]
$\phi=\upsilon$ is the minimum of the scalar field potential. Define the mass
parameter $M$%
\[
M=g\upsilon
\]
$M$ will be regarded as zero order quantity in perturbation. To quantize this
theory, we add to the Lagrangian $L$ gauge fixing terms in the $R_{\xi}$-gauge
as well as the associated ghost terms. The sum will be called the effective
Lagrangian $L_{eff}$ in the $R_{\xi}$-gauge, and is invariant under the
following BRST \cite{BRS,CTBRS} transformation:%
\begin{align}
\delta_{B}A_{\mu}  &  =\partial_{\mu}c,\label{brst}\\
\delta_{B}\phi_{2}  &  =-Mc-gcH,\nonumber\\
\delta_{B}H  &  =gc\phi_{2},\nonumber\\
\delta_{B}\bar{c}  &  =-\dfrac{i}{\xi}\left(  \partial^{\mu}A_{\mu}-\alpha
M\phi_{2}\right)  ,\delta_{B}c=0.\nonumber
\end{align}
where $c$ is the ghost field and $\bar{c}$ is the anti-ghost field. The gauge
fixing term is
\begin{equation}
L_{gf}=-\frac{1}{2\alpha}\left(  \partial_{\mu}A^{\mu}-\alpha M\phi
_{2}\right)  ^{2} \label{egf1}%
\end{equation}
and the ghost term is
\begin{equation}
L_{ghost}=i\bar{c}\delta_{B}\left(  \partial_{\mu}A^{\mu}-\alpha M\phi
_{2}\right)  =i\bar{c}\left(  \partial_{\mu}\partial^{\mu}+\alpha
M^{2}\right)  c+i\alpha Mg\bar{c}Hc. \label{egh1}%
\end{equation}
The effective Lagrangian is%
\begin{equation}
L_{eff}=L+L_{gf}+L_{ghost} \label{e2-3}%
\end{equation}
which gives the following free propagators:%
\[
D\left(  H,H;k\right)  =\frac{i}{k^{2}-\lambda M^{2}};D\left(  \phi_{2}%
,\phi_{2};k\right)  =\frac{i}{k^{2}-\alpha M^{2}}%
\]%
\[
D\left(  A_{\mu},A_{\nu};k\right)  =-\left(  g^{\mu\nu}-\frac{k^{\mu}k^{\nu}%
}{k^{2}}\right)  \frac{i}{\left(  k^{2}-M^{2}\right)  }-\frac{k^{\mu}k^{\nu}%
}{k^{2}}\frac{i\alpha}{k^{2}-\alpha M^{2}}%
\]%
\[
D\left(  \bar{c},c;k\right)  =\frac{1}{k^{2}-\alpha M^{2}};D\left(  c,\bar
{c};k\right)  =-\frac{1}{k^{2}-\alpha M^{2}}%
\]
For the Feynman Gauge, we set $\xi=1$ and obtain%
\[
D\left(  H,H;k\right)  =\frac{i}{k^{2}-\lambda M^{2}};D\left(  \phi_{2}%
,\phi_{2};k\right)  =\frac{i}{k^{2}-M^{2}}%
\]%
\[
D\left(  A_{\mu},A_{\nu};k\right)  =\frac{-ig^{\mu\nu}}{k^{2}-M^{2}}%
\]%
\[
D\left(  \bar{c},c;k\right)  =\frac{1}{k^{2}-M^{2}};D\left(  c,\bar
{c};k\right)  =-\frac{1}{k^{2}-M^{2}}%
\]
Perturbative calculation using the above propagators can be renormalized and
the physical on-shell amplitudes obtained are gauge invariant (independent of
$\xi$). In the limit $\xi\rightarrow\infty$, the propagators become%
\[
D\left(  H,H;k\right)  =\frac{i}{k^{2}-\lambda M^{2}}%
\]%
\begin{align*}
D\left(  A_{\mu},A_{\nu};k\right)   &  =-\left(  g^{\mu\nu}-\frac{k^{\mu
}k^{\nu}}{k^{2}}\right)  \frac{i}{\left(  k^{2}-M^{2}\right)  }+\frac{k^{\mu
}k^{\nu}}{k^{2}}\frac{i}{M^{2}}\\
&  =\frac{-i}{\left(  k^{2}-M^{2}\right)  }\left(  g^{\mu\nu}-\frac{k^{\mu
}k^{\nu}}{M^{2}}\right)
\end{align*}
and all the nonphysical propagators vanish:%
\[
D\left(  \phi_{2},\phi_{2};k\right)  \rightarrow0
\]%
\[
D\left(  \bar{c},c;k\right)  =-D\left(  c,\bar{c};k\right)  \rightarrow0
\]
These $\xi\rightarrow\infty$ propagators are the propagators for the unitary
gauge \cite{UG}. But this does not mean that we disregard all the non-physical
fields in the perturbative calculation of the unitary gauge. This is because
the ghost-Higgs-antighost $c-H-\bar{c}$ vertex factor $-\alpha Mg$ is
proportional to $\xi$ and may annihilate the $1/\xi$ factor due the ghost
propagators in a ghost loop. We shall also make use of the refection symmetry
$\ell_{\sigma}\rightarrow-\ell_{\sigma}$ for any component of the loop
momentum to ignore a Feynman integrand that is odd under $\ell\rightarrow
-\ell,$%
\begin{equation}
\ell_{\mu}f\left(  \ell^{2}\right)  \rightarrow0,\;\ell_{\mu}\ell_{\nu}%
\ell_{\rho}f\left(  \ell^{2}\right)  \rightarrow0,\;\ell_{\mu}\ell_{\nu}%
\ell_{\rho}\ell_{\sigma}\ell_{\tau}f\left(  \ell^{2}\right)  \rightarrow
0,\ldots\label{symodd}%
\end{equation}
and to make the substitutions:%

\begin{equation}
\ell_{\mu}\ell_{\nu}f\left(  \ell^{2}\right)  \rightarrow\frac{g_{\mu\nu}}%
{n}\ell^{2}f\left(  \ell^{2}\right)  \label{symi2}%
\end{equation}
and%
\begin{equation}
\ell_{\mu}\ell_{\nu}\ell_{\rho}\ell_{\sigma}f\left(  \ell^{2}\right)
\rightarrow\frac{g_{\mu\nu}g_{\rho\sigma}+g_{\mu\rho}g_{\nu\sigma}%
+g_{\mu\sigma}g_{\rho\nu}}{n\left(  n+2\right)  }\left(  \ell^{2}\right)
^{2}f\left(  \ell^{2}\right)  , \label{symi4}%
\end{equation}
where $n$ is the space-time dimension. We now proceed to calculate the
integrands for the 2-point $HH$ function given that the external $H$ field is
on mass shell.

\section{1-Loop Self-Energy Correction for Higgs Field}

There are 12 different diagrams for the 2-point $HH$ function. The external
momentum $k$ is on-shell with $k^{2}=\lambda M^{2}$. They are listed below and
categorized into gauge invariant and gauge variant groups. By shifting loop
momentum and using symmetric substitutions as dictated in $\left(
\ref{symodd}\right)  $, $\left(  \ref{symi2}\right)  $ and $\left(
\ref{symi4}\right)  $, these integrands are cast into the forms tabulated in
Tables $\left(  \ref{tb-1}\right)  $ and $\left(  \ref{tb-2}\right)  $.

\paragraph{Gauge Invariant Amplitudes}

\begin{figure}[H]%
\begin{center}
\includegraphics[
height=0.9228in,
width=2.3462in
]%
{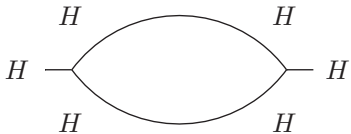}%
\\
{}%
\end{center}
\caption{$-\frac{9}{2}\lambda^{2}g^{2}M^{2}D\left(  H,H;\ell+k\right)
D\left(  H,H;\ell\right)  $}
\label{fighh3}%
\end{figure}

\begin{figure}[H]%
\begin{center}
\includegraphics[
height=0.9245in,
width=1.2894in
]%
{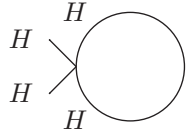}%
\\
{}%
\end{center}

\caption{$-\frac{3}{2}i\lambda g^{2}D\left(  H,H;\ell\right)  $}
\label{fighh2}%
\end{figure}

\begin{figure}[H]%
\begin{center}
\includegraphics[
height=0.9245in,
width=1.5705in
]%
{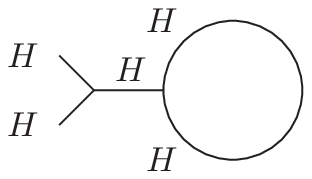}%
\\
{}%
\end{center}
\caption{$-\frac{9}{2}\lambda^{2}g^{2}M^{2}D\left(  H,H,0\right)  D\left(
H,H,\ell\right)  $}
\label{fighh1}%
\end{figure}

Three diagrams that contain only internal $HH$ lines are gauge invariant
because the propagator $D\left(  H,H,k\right)  $ and the 3-poinit $HHH$ vertex
factor and the 4-point $HHHH$ vertex factor are independent of the gauge
parameter $\xi$. These diagrams Figs. $\ref{fighh3}$ - $\ref{fighh1}$ and
their associated integrands are listed in the following.%

\begin{table}[H] \centering
\begin{tabular}
[c]{|l|l|l|}\hline
Diagram & Feynman Gauge & Unitary Gauge\\\hline
Fig. $\ref{fighh3}$ & $\frac{9\lambda^{2}g^{2}M^{2}}{2\left(  \ell^{2}-\lambda
M^{2}\right)  \left(  \left(  \ell+k\right)  ^{2}-\lambda M^{2}\right)  }$ &
$\frac{9\lambda^{2}g^{2}M^{2}}{2\left(  \ell^{2}-\lambda M^{2}\right)  \left(
\left(  \ell+k\right)  ^{2}-\lambda M^{2}\right)  }$\\
Fig. $\ref{fighh2}$ & $\frac{3g^{2}\lambda}{2\left(  \ell^{2}-\lambda
M^{2}\right)  }$ & $\frac{3g^{2}\lambda}{2\left(  \ell^{2}-\lambda
M^{2}\right)  }$\\
Fig. $\ref{fighh1}$ & $-\frac{9\lambda g^{2}}{2\left(  \ell^{2}-\lambda
M^{2}\right)  }$ & $-\frac{9\lambda g^{2}}{2\left(  \ell^{2}-\lambda
M^{2}\right)  }$\\
SUM & $-\frac{3\lambda g^{2}}{\left(  \ell^{2}-\lambda M^{2}\right)  }%
+\frac{9\lambda^{2}g^{2}M^{2}}{2\left(  \ell^{2}-\lambda M^{2}\right)  \left(
\left(  \ell+k\right)  ^{2}-\lambda M^{2}\right)  }$ & $-\frac{3\lambda g^{2}%
}{\left(  \ell^{2}-\lambda M^{2}\right)  }+\frac{9\lambda^{2}g^{2}M^{2}%
}{2\left(  \ell^{2}-\lambda M^{2}\right)  \left(  \left(  \ell+k\right)
^{2}-\lambda M^{2}\right)  }$\\\hline
\end{tabular}
\caption{Gauge Invariant Amplitudes\label{tb-1}}%
\end{table}%

\paragraph{Gauge Variant Amplitudes}

\begin{figure}[H]%
\begin{center}
\includegraphics[
height=0.9219in,
width=2.3462in
]%
{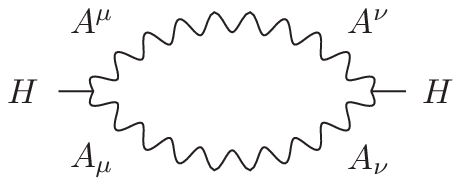}%
\\
{}%
\end{center}
\caption{$-2g^{2}M^{2}D\left(  A^{\mu},A_{\nu};\ell+k\right)  D\left(  A^{\nu
},A_{\mu};\ell\right)  $}
\label{fig21}%
\end{figure}

\begin{figure}[H]%
\begin{center}
\includegraphics[
height=0.9219in,
width=2.3454in
]%
{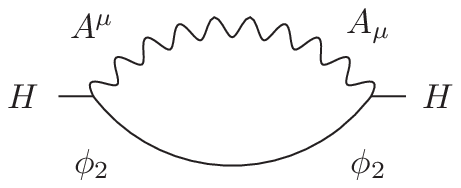}%
\\
{}%
\end{center}
\caption{$-g^{2}D\left(  A^{\mu},A^{\nu};\ell+k\right)  D\left(  \phi_{2}%
,\phi_{2};\ell\right)  \left(  \ell-k\right)  _{\mu}\left(  \ell-k\right)
_{\nu}$}
\label{fig22}%
\end{figure}

\begin{figure}[H]%

\begin{center}
\includegraphics[
height=0.9228in,
width=2.3454in
]%
{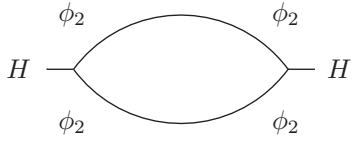}%
\\
{}%
\end{center}
\caption{$-\frac{1}{2}\lambda^{2}g^{2}M^{2}D\left(  \phi_{2},\phi_{2}%
;\ell+k\right)  D\left(  \phi_{2},\phi_{2};\ell\right)  $}
\label{fig25}%
\end{figure}

\begin{figure}[H]%

\begin{center}
\includegraphics[
height=0.9219in,
width=2.3462in
]%
{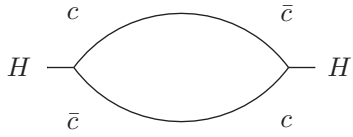}%
\\
{}%
\end{center}
\caption{$-\alpha^{2}g^{2}M^{2}D\left(  c,\bar{c},\ell\right)  D\left(
c,\bar{c},\ell+k\right)  $}
\label{fig26}%
\end{figure}

\begin{figure}[H]%

\begin{center}
\includegraphics[
height=0.9236in,
width=1.3474in
]%
{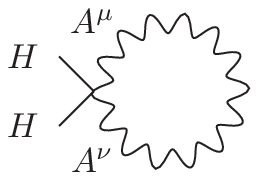}%
\\
{}%
\end{center}
\caption{$ig^{2}D\left(  A^{\mu},A_{\mu};\ell\right)  $}
\label{fig23}%
\end{figure}

\begin{figure}[H]%

\begin{center}
\includegraphics[
height=0.9236in,
width=1.2894in
]%
{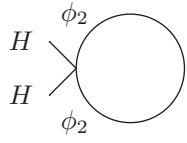}%
\\
{}%
\end{center}
\caption{$-\frac{1}{2}ig^{2}\lambda D\left(  \phi_{2},\phi_{2};\ell\right)  $}
\label{fig24}%
\end{figure}

\begin{figure}[H]%

\begin{center}
\includegraphics[
height=0.9236in,
width=1.6129in
]%
{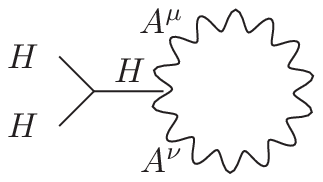}%
\\
{}%
\end{center}
\caption{$3\lambda g^{2}M^{2}D\left(  H,H,0\right)  D\left(  A^{\mu},A_{\mu
},\ell\right)  $}
\label{fig29}%
\end{figure}

\begin{figure}[H]%

\begin{center}
\includegraphics[
height=0.9236in,
width=1.5705in
]%
{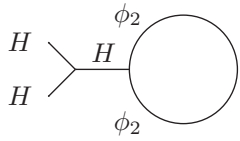}%
\\
{}%
\end{center}
\caption{$-\frac{3}{2}\lambda^{2}g^{2}M^{2}D\left(  H,H,0\right)  D\left(
\phi_{2},\phi_{2},\ell\right)  $}
\label{fig28}%
\end{figure}

\begin{figure}[H]%

\begin{center}
\includegraphics[
height=0.9236in,
width=1.5705in
]%
{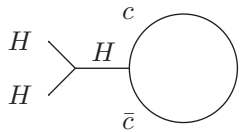}%
\\
{}%
\end{center}
\caption{$-3i\alpha\lambda g^{2}M^{2}D\left(  H,H,0\right)  D\left(  c,\bar
{c},\ell\right)  $}
\label{fig27}%
\end{figure}

Figs. $\ref{fig21}$ - $\ref{fig27}$ are the diagrams that contain non-physical
internal lines. Although the two ghost propagators in Figure $\ref{fig26}$
vanishes (like $1/\xi$) as $\xi\rightarrow\infty$ in the unitary gauge, there
are two vertex factors each of which is proportional to $\xi$. The integrand
for Figure $\ref{fig26}$ therefore does not vanish in the unitary gauge.
Similarly, the tadpole diagram of Figure $\ref{fig27}$ with one ghost
propagator cannot be ignored in the unitary gauge.%

\begin{table}[H] \centering
\begin{tabular}
[c]{lll}%
Diagram & Feynman Gauge & Unitary Gauge\\
Fig. $\ref{fig21}$ & $\frac{2ng^{2}M^{2}}{\left(  \ell^{2}-M^{2}\right)
\left(  \left(  \ell+k\right)  ^{2}-M^{2}\right)  }$ & $\frac{2g^{2}}{M^{2}%
}-\frac{g^{2}\lambda}{\ell^{2}-M^{2}}+\frac{g^{2}M^{2}\left[  2\left(
n-1\right)  +\frac{\lambda^{2}}{2}-2\lambda\right]  }{\left(  \ell^{2}%
-M^{2}\right)  \left(  \left(  \ell+k\right)  ^{2}-M^{2}\right)  }$\\
Fig. $\ref{fig22}$ & $-\frac{g^{2}}{\ell^{2}-M^{2}}-\frac{g^{2}M^{2}\left(
1+2\lambda\right)  }{\left(  \ell^{2}-M^{2}\right)  \left(  \left(
\ell+k\right)  ^{2}-M^{2}\right)  }$ & $0$\\
Fig. $\ref{fig25}$ & $\frac{\lambda^{2}g^{2}M^{2}}{2\left(  \ell^{2}%
-M^{2}\right)  \left(  \left(  \ell+k\right)  ^{2}-M^{2}\right)  }$ & $0$\\
Fig. $\ref{fig26}$ & $-\frac{g^{2}M^{2}}{\left(  \ell^{2}-M^{2}\right)
\left(  \left(  \ell+k\right)  ^{2}-M^{2}\right)  }$ & $-\frac{g^{2}}{M^{2}}%
$\\
Fig. $\ref{fig23}$ & $\frac{ng^{2}}{\left(  \ell^{2}-M^{2}\right)  }$ &
$-\frac{g^{2}}{M^{2}}+\frac{\left(  n-1\right)  g^{2}}{\left(  \ell^{2}%
-M^{2}\right)  }$\\
Fig. $\ref{fig24}$ & $\frac{g^{2}\lambda}{2\left(  \ell^{2}-M^{2}\right)  }$ &
$0$\\
Fig. $\ref{fig29}$ & $-\frac{3ng^{2}}{\left(  \ell^{2}-M^{2}\right)  }$ &
$\frac{3g^{2}}{M^{2}}-\frac{3\left(  n-1\right)  g^{2}}{\left(  \ell^{2}%
-M^{2}\right)  }$\\
Fig. $\ref{fig28}$ & $-\frac{3\lambda g^{2}}{2\left(  \ell^{2}-M^{2}\right)
}$ & $0$\\
Fig. $\ref{fig27}$ & $\frac{3g^{2}}{\left(  \ell^{2}-M^{2}\right)  }$ &
$-\frac{3g^{2}}{M^{2}}$\\
SUM & $\frac{g^{2}\left[  2-2n-\lambda\right]  }{\ell^{2}-M^{2}}+\frac
{g^{2}M^{2}\left[  2n-2-2\lambda+\frac{\lambda^{2}}{2}\right]  }{\left(
\ell^{2}-M^{2}\right)  \left(  \left(  \ell+k\right)  ^{2}-M^{2}\right)  }$ &
$\frac{g^{2}\left[  2-2n-\lambda\right]  }{\ell^{2}-M^{2}}+\frac{g^{2}%
M^{2}\left[  2n-2-2\lambda+\frac{\lambda^{2}}{2}\right]  }{\left(  \ell
^{2}-M^{2}\right)  \left(  \left(  \ell+k\right)  ^{2}-M^{2}\right)  }$%
\end{tabular}
\caption{Gauge Variant Amplitudes\label{tb-2}}%
\end{table}%

\section{Conclusion}

The last row of Table \ref{tb-2} indicates that the sum of all gauge variant
integrands for the Feynman gauge is equal to that for the unitary gauge
provided dimensional regularization is utilized in the manipulation of the
Feynman integrand of every individual diagram.

In the diagrammatic proof of gauge invariance for on-shell amplitude, all we
need is making shifts or refections of loop variables on the integrands of
Feynman diagrams. Dimensional regularization ensures the legitimacy to shift
or reflect loop momenta and hence guarantees gauge symmetry. The sum of
integrands for the alpha gauge should be $\xi$ independent in the scheme of
dimensional gauge. Without the need of carrying out the actual integration of
Feynman integrals, the labor involved in verifying gauge invariance is greatly
reduced and enables us to demonstrate the existence of ghost loop in the
unitary gauge.

\end{document}